\documentclass[aps,prb,showpacs,superscriptaddress,twocolumn]{revtex4}
\usepackage{graphicx}
\usepackage{amsmath}
\usepackage{bm}
\usepackage{xcolor}

\begin{document}

\title{Influence of magnetic disorders on Quantum Anomalous Hall Effect in Magnetic Topological Insulator Films beyond the two-dimensional limit}

\author{Yanxia Xing}
\email[]{xingyanxia@bit.edu.cn}
\affiliation{Beijing Key Laboratory of Nanophotonics and Ultrafine Optoelectronic Systems, School of Physics, Beijing Institute of Technology, Beijing 100081, China}

\author{King Tai Cheung}
\affiliation{Department of Physics and the Center of Theoretical and Computational Physics, The University of Hong Kong, Pokfulam Road, Hong Kong, China}

\author{Fuming Xu}
\affiliation{College of Physics and Energy, Shenzhen University, Shenzhen 518060, China}

\author{Qing-Feng Sun}
\affiliation{International Center for Quantum Materials, School of Physics, Peking University, Beijing 100871, China}
\affiliation{Collaborative Innovation Center of Quantum Matter, Beijing 100871, China}

\author{Jian Wang}
\affiliation{Department of Physics and the Center of Theoretical and Computational Physics, The University of Hong Kong, Pokfulam Road, Hong Kong, China}

\author{Yugui Yao}
\affiliation{Beijing Key Laboratory of Nanophotonics and Ultrafine Optoelectronic Systems, School of Physics, Beijing Institute of Technology, Beijing 100081, China}

\begin{abstract}
Quantum anomalous Hall effect (QAHE) has been experimentally realized in magnetic topological insulator (MTI) thin films fabricated
on magnetically doped $(Bi,Sb)_2Te_3$. In a MTI thin film with the magnetic easy axis along the normal direction (z-direction),
orientations of magnetic dopants are randomly distributed around the magnetic easy axis, acting as magnetic disorders.
With the aid of the non-equilibrium Green's function and Landauer-B$\ddot{u}$ttiker formalism, we numerically study the influence of
magnetic disorders on QAHE in a MTI thin film modeled by a three-dimensional tight-binding Hamiltonian. It is found that, due to the
existence of gapless side surface states, QAHE is protected even in the presence of magnetic disorders as long as the z-component of
magnetic moment of all magnetic dopants are positive. More importantly, such magnetic disorders also suppress the dissipation of the chiral edge states and enhance the quality of QAHE in MTI films. In addition, the effect of magnetic disorders depends very much on the film thickness, and the optimal influence is achieved at certain thickness. These findings are new features for QAHE in three-dimensional systems, not present in two-dimensional systems.
\end{abstract}



\maketitle

\section{introduction}

The quantum anomalous Hall effect (QAHE) is of interest to both fundamental research and spintronic applications.\cite{PhysicalReviewB68.045327.Culcer.2003,PhysicalReviewLetters92.037204.Yao.2004,ProceedingsoftheRoyalSocietyA:MathematicalPhysicalandEngineeringSciences392.45-57.Berry.1984,Rev.Mod.Phys.82.1959.Xiao.2010} QAHE was originally proposed in various ideal 2-dimensional(2D) systems, including 2D honeycomb lattices with periodic pseudo-magnetic fields,\cite{Phys.Rev.Lett.61.2015.Haldane.1988,Shen.2012} MnHgTe magnetic quantum wells,\cite{PhysicalReviewLetters101.146802.Liu.2008} monolayer\cite{PhysicalReviewB82.Qiao2010} or bilayer\cite{PhysicalReviewB83.Tse2011} graphenes, and multilayer topological insulators with magnetic doping.\cite{PhysicalReviewB85.045445.Jiang.2012,PhysicalReviewLetters111.136801.Wang.2013}

After the theoretical prediction of QAHE in three-dimensional(3D) magnetic topological insulator (MTI) thin films,\cite{Science329.61-64.Yu.2010} the existence of QAHE was recently verified by a series of experiments in magnetically doped $(Bi,Sb)_2Te_3$ systems.\cite{Science340.167-170.Chang.2013,PhysicalReviewLetters113.Kou.2014,PhysicalReviewLetters114.187201.Bestwick.2015,NatureMaterials14.473-477.Chang.2015}
Different from the conventional QAHE discussed in 2D cases,\cite{Phys.Rev.Lett.61.2015.Haldane.1988,PhysicalReviewLetters101.146802.Liu.2008,PhysicalReviewB82.Qiao2010,PhysicalReviewB83.Tse2011,PhysicalReviewB85.045445.Jiang.2012,PhysicalReviewLetters111.136801.Wang.2013}
the quantized Hall conductance in 3D MTI films is jointly contributed by the top and bottom massive Dirac-like surface states which have opposite signs in their effective masses.\cite{Science329.61-64.Yu.2010,Phys.Rev.B84.085312.Chu.2011} However, the gapless side surfaces are still crucial for QAHE in 3D MTI films. Since the top and bottom surfaces are gapped, the chiral edge modes actually propagate through the gapless side surface states. Besides the side surface states, a constant exchange field $\mathbf{M}$ is essential for QAHE in MTI films, responsible for breaking the time-reversal symmetry and consequently opening the nontrivial energy surface gap. Experimentally, it is inevitable to have disorders present in the system especially for the magnetically doped system $(Bi,Sb)_2Te_3$.
However, it is very difficult to align all magnetic moments of magnetic dopant experimentally. The random deviation of magnetic moments away from easy axis of magnetization along z-direction is a source of magnetic disorders.
It would be interesting to investigate the effect of magnetic disorders on QAHE in a 3D MTI film.

Generally speaking, both external magnetic fields\cite{PhysicalReviewBAvishai1995} and local magnetic moments \cite{PhysicalReviewBean1962,PhysicalReviewBTakahashi1999,PhysicalReviewBIvanov2009} can induce the spin dependent energy split for  electrons. In this paper, we consider only the later. Here, the local magnetic moments originates from the magnetic dopants.
In the experiments of QAHE in MTI films such as $(Bi,Sb)_2Te_3$ film, magnetic doping gives rise to a macroscopic magnetism along z direction (perpendicular to the film) needed to produce QAHE. However, it is very difficult to align all magnetic moments of magnetic dopant experimentally. The random deviation of magnetic moments away from macroscopic magnetization along z-direction is a type of magnetic disorders considered in this paper.
\begin{figure}
\includegraphics[width=8.5cm, clip=]{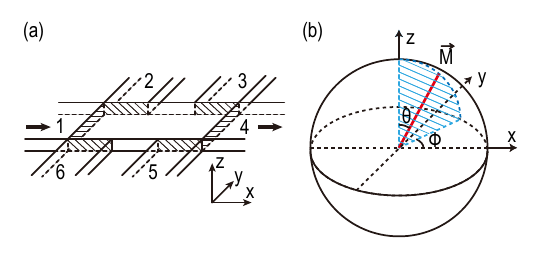}
\caption{ (Color online)
Panel (a): schematic plot of a six-terminal Hall device. The current is injected from terminal-1 to terminal-4.
Panel (b): angular representation of $\mathbf{M}$.}\label{fig01}
\end{figure}

In this paper, with the aid of the non-equilibrium Green's functions and Landauer-B$\ddot{u}$ttiker formula, we study the influence of magnetic disorders on QAHE in a six-terminal Hall bar constructed on a MTI film, which is illustrated in Fig.\ref{fig01}(a). In the calculation, the orientation of exchange field $\mathbf{M}$ is represented by $(\theta,\phi)$, as shown in Fig.\ref{fig01}(b). In this representation, $\mathbf{M}_z=|\mathbf{M}|\cos\theta$, $\mathbf{M}_{xy}=|\mathbf{M}|\sin\theta$, where $\mathbf{M}_z$ is the $z$-component of the exchange field while $\mathbf{M}_{xy}$ is its projection in the $x$-$y$ plane. To include contributions of side surface states, we adopt the 3D Hamiltonian\cite{Phys.Rev.B89.245107.Zhang.2014} derived from bulk $(Bi,Sb)_2Te_3$ material instead of the previous 2D effective Hamiltonian.\cite{Science329.61-64.Yu.2010,PhysicalReviewLetters111.146802.Lu.2013,Phys.Rev.B89.155419.Zhang.2014} Our numerical results suggest that, in the presence of random magnetic disorders, QAHE is always robust as long as $\mathbf{M}_z>0$ for all magnetic dopants, regardless of the $x$-$y$ component $\mathbf{M}_{xy}$. More importantly, the random distribution of $\mathbf{M}$ can suppress the dissipation of the chiral edge states. It should be emphasized that these conclusions are valid only in 3D systems, where the side surface in the 3D thin film plays a crucial role. In fact, 3D QAHE relies much on the thickness $C_z$ of the MTI film and is most sensitive to the distribution of $\mathbf{M}$ at the film thickness of $C_z\simeq 5nm$.
Since magnetic disorders are inevitable in three-dimensional MTI films, these findings can serve as an useful guide for the application of QAHE.

The rest of the paper is organized as follows. Using a four-band tight-binding model, the Hamiltonian of the six-terminal Hall system is introduced in Sec. II. The formalisms for calculating the longitudinal resistance and Hall resistance are also derived. Sec. III gives numerical results of the effect of magnetic disorders on QAHE in the 3D MTI film, accompanied with detailed discussions. Finally, a brief summary is presented in Sec. IV.

\section{model and theory}

Using $k\cdot p$ perturbation theory, the low energy spectrum of $(Bi,Sb)_2Te_3$ can be approximated by
the four-band Hamiltonian $H_0(k)$ which is written as\cite{Phys.Rev.BLiu2010,NatPhys5.438-442.Zhang.2009}
\begin{equation}
H_0(k)=\epsilon_k  + \mathbf{p}\cdot \mathbf{\sigma}\otimes\tau_x + m_k \sigma_z\otimes\tau_0\label{H0},
\end{equation}
with $\mathbf{p}=[A_xk_x,A_yk_y,A_z k_z]$ and $m_k=m_0-m_xk_x^2-m_yk_y^2-m_z k_z^2$.
Eq.(\ref{H0}) is the Dirac equation of 3D systems, consists of four-component Dirac matrices $\sigma_{i=x,y,z}\otimes\tau_x$ and $\sigma_z\otimes\tau_0$, where $\sigma$ and $\tau$ are Pauli matrices for spin ($\uparrow/\downarrow$) and orbit ($P_{1z}^+/P_{2z}^-$), respectively. $\sigma_0$ and $\tau_0$ are $2\times 2$ unit matrices.
For simplicity, we set $\epsilon_k=0$ in the calculation since it doesn't change the topological structure of the Hamiltonian. To study the magnetic doping effect, we express the Hamiltonian as
\begin{equation}
H(k)=H_0(k)+\mathbf{M}\cdot\sigma\otimes\tau_0\label{H1},
\end{equation}
where $\mathbf{M}$ is the exchange field induced by magnetic doping. To describe a multi-terminal device fabricated on magnetically doped $(Bi,Sb)_2Te_3$ films,
a real-space Hamiltonian is needed. Replacing $k_{x,y,z}$ by $-i\nabla_{x,y,z}$, we get the 3D effective tight-binding Hamiltonian on a cubic lattice:\cite{Phys.Rev.B89.245107.Zhang.2014}
\begin{equation}
H = \sum_{\mathbf{i}}d_{\mathbf{i}}^{\dag}H_{\mathbf{i}}d_{\mathbf{i}}
+\sum_{\mathbf{i},\delta}\left[d_{\mathbf{i}}^{\dag}H_{\delta} d_{\mathbf{i}+\delta}
+d_{\mathbf{i}+\delta}^{\dag}H_{\delta}^\dagger d_{\mathbf{i}}\right], \label{HReal}
\end{equation}
where $\mathbf{i}=[\mathbf{i}_x,\mathbf{i}_y,\mathbf{i}_z]$ labels the site in real space, $\delta=\mathbf{a}_\alpha$ with $\alpha=x,y,z$ and $\mathbf{a}_\alpha$ are basis vectors of the cubic lattice, $H_\mathbf{i}$ describes the on-site potential, while $H_\delta$ and $H_\delta^\dagger$ denotes hopping to the six nearest neighbors in the cubic lattice. Their expressions are given by, respectively
\begin{equation}
\begin{split}
  &H_{\mathbf{i}} =\mathbf{M_i}\cdot\sigma\tau_{0}
  +(m_{0}+2\sum_{\alpha}\frac{m_{\alpha}}{a^2})\sigma_{0}\tau_{z},\\
  &H_\delta=H_{\mathbf{a}_\alpha}=\left[-\frac{M_{\alpha}}{a^2}\sigma_{0}\tau_{z}-i \frac{A_{\alpha}}{2a}\sigma_{\alpha}\tau_{x}\right]e^{i\phi_{\mathbf{i},\mathbf{i}+\mathbf{a}_\alpha}}\label{HReal1},
  \end{split}
\end{equation}
where $\mathbf{M_i}$ is the random exchange field at site i.
Since orientations of the local magnetic moments randomly distribute, we assume that the angle $(\theta,\phi)$ [see Fig.1(b)] of $\mathbf{M_i}$ is random, while its magnitude $|\mathbf{M_i}|$ is a constant.
In Eq.(\ref{HReal1}), $a=|\mathbf{a}|$ is the lattice constant of the cubic lattice.
In the presence of a perpendicular magnetic field $\mathbf{B}=[0,0,B]$, an extra phase $\phi_{\mathbf{i},\mathbf{i}+a_{\alpha}}=\frac{e}{\hbar}\int_{\mathbf{i}}^{\mathbf{i}+a_{\alpha}}\mathbf{A}\cdot \mathbf{dl}$ is induced in the nearest neighbor coupling term $H_\delta$ with $\mathbf{B}=\nabla\times\mathbf{A}$.

In order to calculate the Hall resistance $\rho_{xy}$ and longitudinal resistance $\rho_{xx}$, we consider the six-terminal Hall bar system shown in Fig.1(a). In this setup, the central scattering region is connected to six semi-infinite leads. In the Coulomb gauge, the vector potential is chosen as $\mathbf{A}=[-By,0,0]$ in lead-1, lead-4, and the central scattering region, which does not depend on x. For lead-2, lead-3, lead-5 and lead-6  $\mathbf{A}=[0,Bx,0]$ is used  which is independent of $y$. The magnetic flux in each unit cell in x-y plane satisfies $\Phi_0=Ba^2$ for every layer in the $x$-$y$ plane since the magnetic field $B$ is constant. Since two different gauges are used for different regions, we have to be careful in maintaining constant magnetic flux at the boundaries between scattering region and leads. This can be done through gauge transformation in the boundaries between the scattering region and lead-2, lead-3, lead-5 and lead-6. In the following calculation, other parameters in Eq.(\ref{HReal}) and Eq.(\ref{HReal1}) are set as $a=5$\AA, $|\mathbf{M}|=0.15$eV, $m_{0}=0.28$eV,\cite{Phys.Rev.BLiu2010} $m_{x}=m_{y}=56.6$eV{\AA}$^2$, $m_{z}=10$eV{\AA}$^2$, $A_{z}=2.2$eV{\AA} and $A_x=A_y=4.1$eV{\AA}, respectively.\cite{NatPhys5.438-442.Zhang.2009}

The current from the $m$-th lead can be calculated using Landau-B\"{u}ttiker formalism,\cite{S.Datta.1995} which expresses as
\begin{equation}
J_m = \frac{e^2}{h}\sum_{n}T_{mn}(V_m-V_n)\label{LB},
\end{equation}
where $m,n=1,2,...,6$ label the leads, and $T_{mn}$ is the transmission coefficient from lead $n$ to lead $m$. The transmission coefficient is expressed as $T_{mn}={\rm Tr}[\Gamma_mG^r\Gamma_nG^a]$, where "${\rm Tr}$" denotes the trace. The line width function $\Gamma_m$ is defined as $\Gamma_{m} = i[\Sigma^r_{m}-\Sigma^{a}_{m}]$, and $G^r$ is the retarded Green's function $G^r = [G^{a}]^\dagger=(EI-H_{c}-\sum_m \Sigma^{r}_{m})^{-1}$, where $H_c$ is the Hamiltonian of the central scattering region. $I$ is an unit matrix with the same dimension of $H_{c}$, and $\Sigma^{r}_{m}$ is the retarded self energy contributed by the semi infinite lead-$m$ which can be obtained using the transfer-matrix method.\cite{Phys.Rev.B23.4988.1981,Phys.Rev.B23.4997.1981}

In the measurement of QAHE, a bias voltage is applied across terminal-1 and terminal-4 to drive the current. The other four terminals serve as voltage probes. By requiring $J_{2,3,5,6}=0$, voltages $V_{2,3,5,6}$ can be determined. With these voltages, together with $V_1 = V$ and $V_4 = 0$, we can calculate the current $J_1 = -J_4$ from Eq.(\ref{LB}). Finally, the longitudinal resistance $\rho_{xx} \equiv (V_2-V_3)/J_1$ and Hall resistance $\rho_{xy}\equiv (V_6-V_2)/J_1$ are obtained. For a perfect Hall effect, $\rho_{xx}$ is exactly zero and $\rho_{xy}$ is ideally quantized.

\section{numerical results and discussions}

For QAHE in a MTI film with finite thickness, the dimension along z which is perpendicular to the film is important for two reasons. First of all, the top and bottom surface states are coupled to each other in $z$-direction. This coupling results in an effective mass term which in turn leads to the band inversion. Consequently this gives rise to the topological transition, i.e., QAHE, in the simplest low-energy effective Hamiltonian consisting of Dirac-type surface states only.\cite{Science329.61-64.Yu.2010,PhysicalReviewLetters111.146802.Lu.2013,Phys.Rev.B89.155419.Zhang.2014}
Secondly, in the presence of exchange field $\mathbf{M}$ along $z$-direction, the top and bottom surface states are gapped while the side surfaces are gapless. As a result, the chiral edge states of the bottom and top surfaces are actually propagating through gapless side surfaces. Therefore, it is necessary to study the influence of the film thickness on QAHE.

\begin{figure}
\includegraphics[width=7cm,totalheight=6.5cm, clip=]{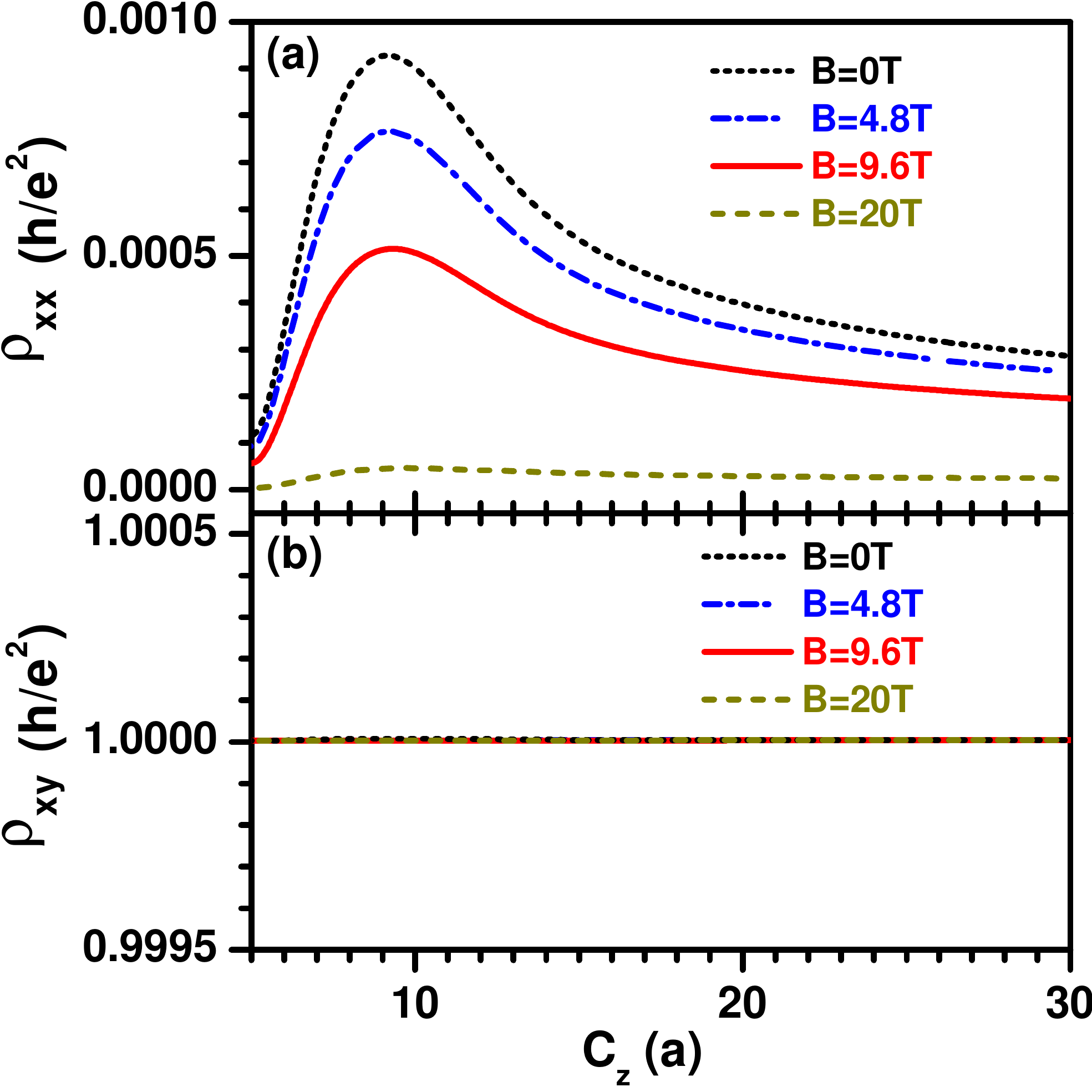}
\caption{ (Color online)
Influence of the film thickness $C_z$. (a) and (b) $\rho_{xx}$ and $\rho_{xy}$ vs $C_z$ for different magnetic field strengths. Here $E_F=0.01$eV}
\end{figure}

To begin with, we plot in Fig.2 the longitudinal resistance $\rho_{xx}$ and the Hall resistance $\rho_{xy}$ against thickness $C_z$ of the MTI film for different magnetic field strengths $B$ in the presence of constant exchange field, i.e., $M_z=0.15$eV and $M_{xy}=0$. Here, a magnetic field is also applied to suppress the dissipation and thereby improve the quality of chiral edge states of QAHE. As expected, the Hall resistance $\rho_{xy}$ is quantized when Fermi energy $E_F=0.01$eV is in the surface gap.
Compared with the quality of Hall resistance, however, the longitudinal resistance $\rho_{xx}$ is not so ideal which shows significant deviation from zero even in the presence of a large magnetic field. From Fig.2 we find that with the increasing of film thickness $C_z$, $\rho_{xx}$ increases quickly and reaches the maximum at $C_z\simeq 10a$, regardless of the magnetic field strength. The stronger the magnetic field, the smaller the longitudinal resistance. Apparently, external magnetic fields indeed improve the quality of QAHE. Therefore, for a constant exchange field, the edge states are most dissipative at the thickness of $C_z=10a$.
\begin{figure*}
\includegraphics[width=16cm, clip=]{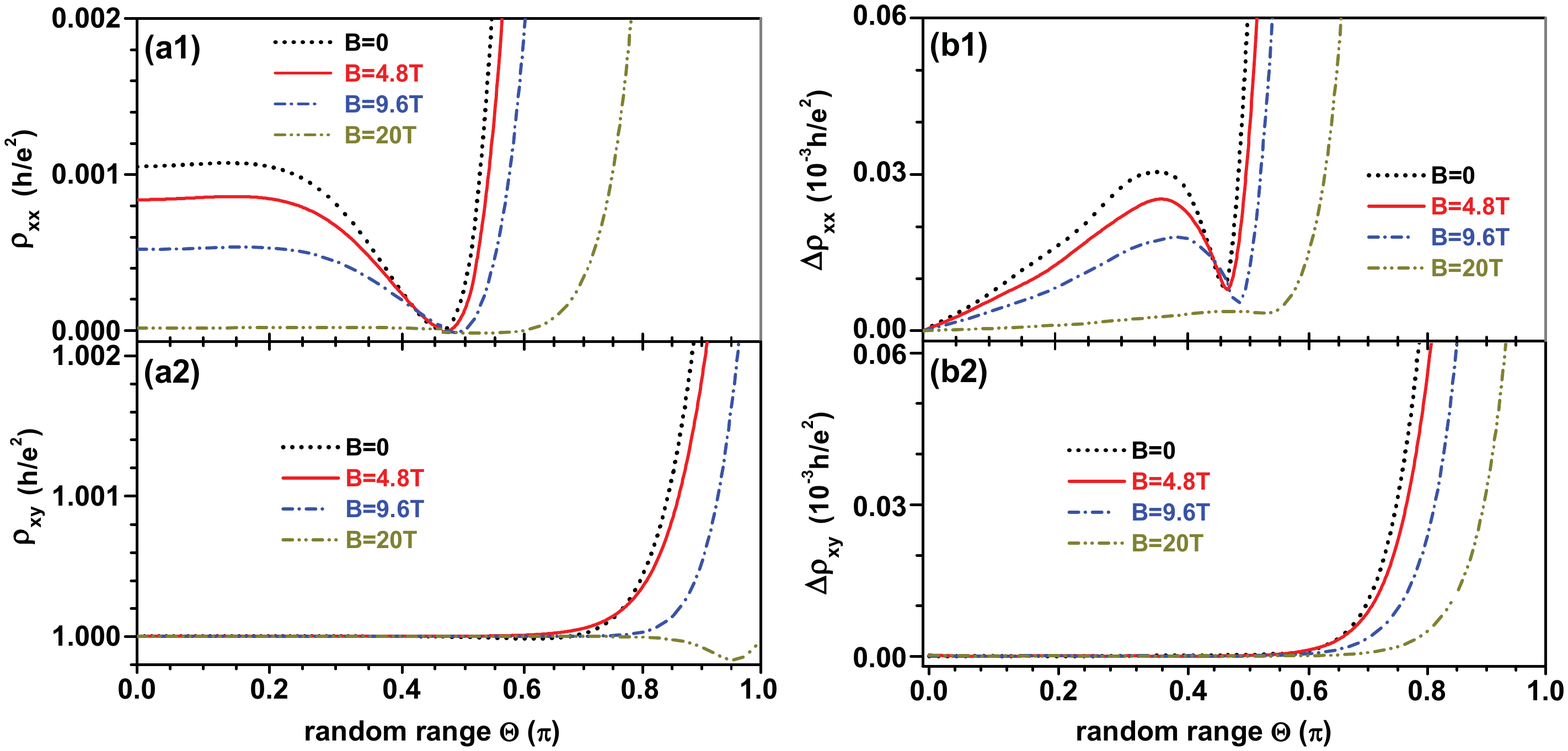}
\caption{ (Color online) $\rho_{xx}$, $\rho_{xy}$ and their fluctuations $\Delta\rho_{xx}$, $\Delta\rho_{xy}$ vs $\Theta$ at fixed film thickness $C_z=10a$. Other parameters: $C_x=90a$, $C_y=30a$, $E_F=0.01$eV. Averaged over 1000 random configurations at each $\Theta$.}
\end{figure*}

\begin{figure}
\includegraphics[width=8.5cm,totalheight=6cm, clip=]{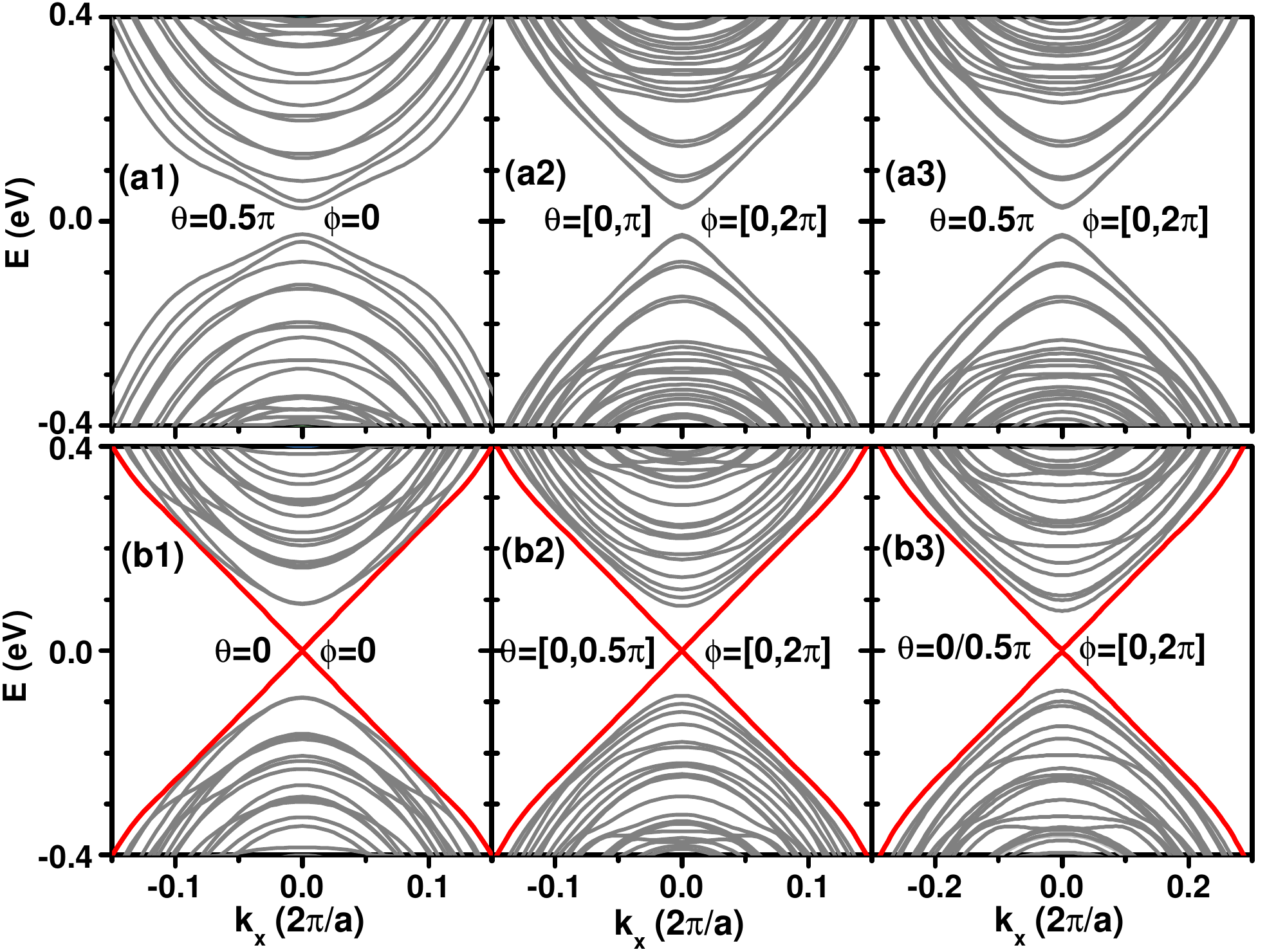}
\caption{ (Color online)
Band structures of 3D MTI thin films with various exchange fields $\mathbf{M}$.
Red lines denotes the chiral edge states.
Panel (a1) the orientation $\mathbf{M}$ is fixed in $x$-direction: $\mathbf{M}=\mathbf{M}_x$;
Panel (a2) the orientation of $\mathbf{M}$ uniformly distributes in the whole $4\pi$ solid angle: $\theta=[0,\pi]$, $\phi=[0,2\pi]$;
Panel (a3) $\mathbf{M}$ is located in the $x$-$y$ plane, and the orientation of $\mathbf{M}$ uniformly distributes in the whole x-y plane: $\mathbf{M}_z=0$, $\phi=[0,2\pi]$;
Panel (b1) the direction of $\mathbf{M}$ is fixed in $z$-direction: $\mathbf{M}=\mathbf{M}_z$;
Panel (b2) $\mathbf{M}$ is uniformly distributed in the upper hemisphere: $\Theta=0.5\pi$, $\phi=[0,2\pi]$;
Panel (b3) the hybrid of $\mathbf{M}=\mathbf{M}_{xy}$ and $\mathbf{M}=\mathbf{M}_z$. }
\end{figure}

Next, at fixed film thickness $C_z=10a$ where the edge states are most dissipative in the clean sample, we allow the direction of local exchange field to fluctuate and study the influence of the random fluctuation on QAHE. Specifically, we denote the orientation of exchange field $\mathbf{M}$ as ($\theta,\phi$) as shown in Fig.1(b). Since the macroscopic magnetization direction of the MTI film is along $z$-direction, we assume that $\phi$ and $\theta$ are in the range of $[0,2\pi]$ and $[0,\Theta]$, respectively, where $\Theta$ measures the largest angular deviation from z-direction.
In Fig.3, we show the average longitudinal resistance $\rho_{xx}$, the average Hall resistance $\rho_{xy}$ and their fluctuations against $\Theta$ at different magnetic fields. From Fig.3(a1) and Fig.3(a2), we find that when $\Theta$ is small ($\Theta<0.2\pi$) $\rho_{xx}$ obviously deviates from zero. Upon increasing $\Theta$ further, $\rho_{xx}$ starts to decrease and reaches a minimum at $\Theta \approx 0.46$-$0.5\pi$, where $\rho_{xx}\approx 0$ for any magnetic field strength.
When $\Theta > 0.5\pi$, $\rho_{xx}$ increases abruptly with increasing of $\Theta$. At the same time, $\rho_{xy}$ is always perfectly quantized until $\Theta>0.6\pi$. When $\Theta > 0.6\pi$ $\rho_{xy}$ starts to deviate from the quantized value abruptly.
Besides $\rho_{xx}$ and $\rho_{xy}$, we also plot their fluctuations $\Delta\rho_{xx}$ and $\Delta\rho_{xy}$ in Fig.3(b1) and Fig.3(b2), which are defined as the root of mean square of the corresponding resistances.
It is found that when $\Theta=0$, both $\Delta\rho_{xx}$ and $\Delta\rho_{xy}$ are zero, and $\rho_{xx}$ and $\rho_{xy}$ don't fluctuate since there is no randomness. With the increasing of $\Theta$, $\Delta\rho_{xx}$ increases to its local maximum and then drops to a local minimum at $\Theta \approx 0.46$-$0.5\pi$. When $\Theta>0.5\pi$, $\Delta\rho_{xx}$ increases abruptly.
On the other hand, when $\Theta<0.6\pi$ the fluctuation of $\rho_{xy}$ is very small compared with $\rho_{xx}$. It starts to increase drastically for $\Theta>0.6\pi$.
Notice that, at $\Theta\approx0.46$-$0.5\pi$, the direction of exchange field $\mathbf{M}$ is randomly distributed almost in the entire upper hemi-sphere, suggesting that QAHE in a MTI film is very robust in the presence of magnetic disorders, as long as all $\mathbf{M}_z > 0$.
More importantly, the moderate angular randomness of magnetic dopant can lead to the suppression of dissipation and improve the quality of chiral edge sates, since both $\rho_{xx}$ and its fluctuation $\Delta\rho_{xx}$ drop to the minimum at $\Theta \approx 0.46$-$0.5\pi$. Furthermore, external magnetic field $B$ obviously suppresses $\rho_{xx}$ but hardly affects $\rho_{xy}$ when $\Theta<0.5\pi$, which is consistent with previous experimental observations.\cite{Science340.167-170.Chang.2013,PhysicalReviewLetters113.Kou.2014}

\begin{figure*}[t]
\includegraphics[width=14cm, clip=]{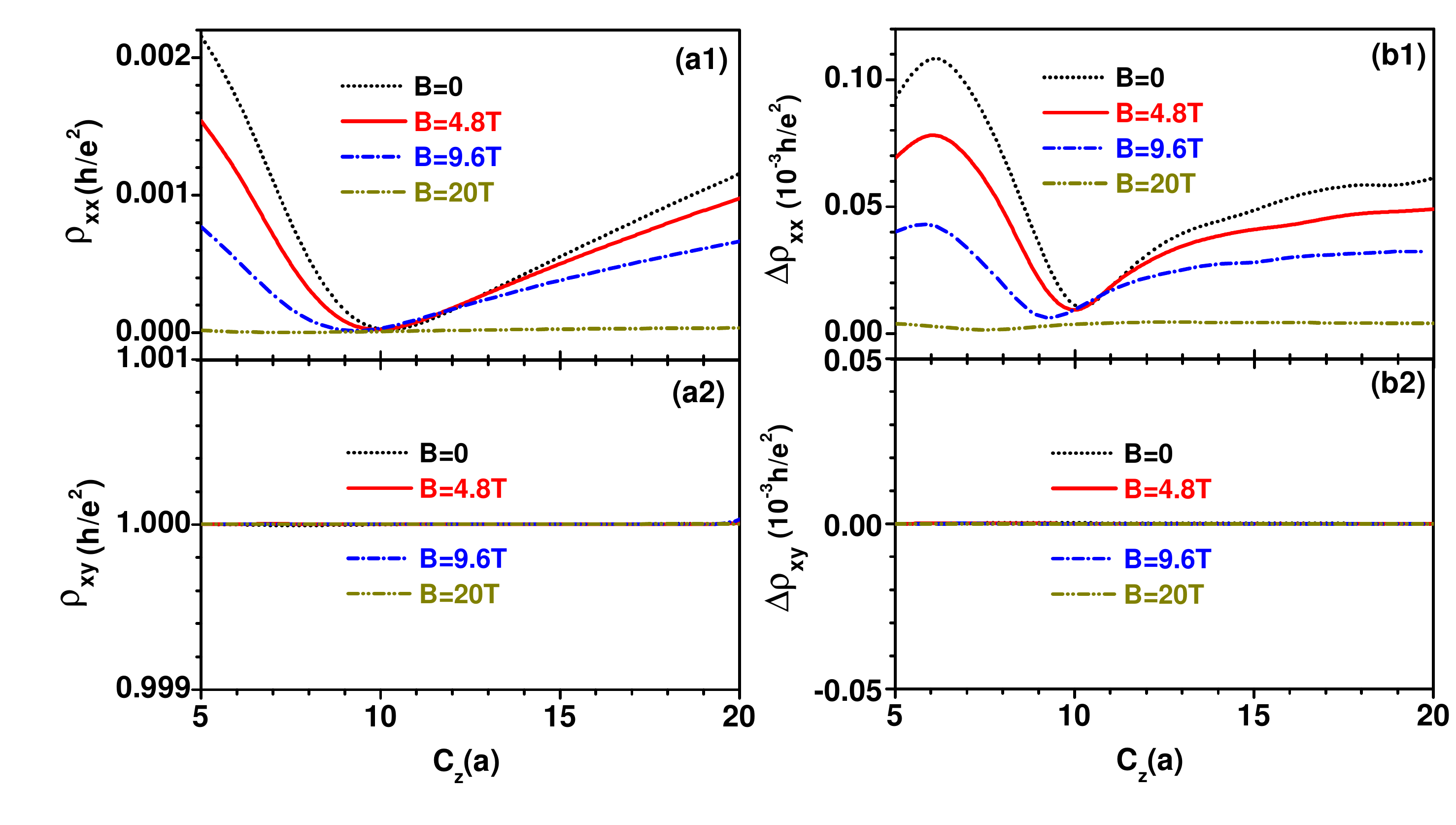}
\caption{ (Color online) Influence of the film thickness $C_z$. $\rho_{xx}$, $\rho_{xy}$ and their fluctuations $\Delta\rho_{xx}$, $\Delta\rho_{xy}$
vs $C_z$ for different magnetic field strengths. The range of random orientation of the exchange field is set as $\theta=[0,0.46\pi]$, $\phi=[0,2\pi]$. Other parameters: $C_x=90a$, $C_y=30a$, $E_F=0.01eV$. Averaged over 1000 random configurations.}
\end{figure*}

The numerical results in Fig.3 is really counterintuitive, in which a moderate angular randomness does not suppress QAHE but enhance it at $\Theta \approx 0.46$-$0.5\pi$. In order to understand this phenomenon, in Fig.4 we show energy bands of 3D MTI thin films with various angular randomness of $\mathbf{M}$. In the presence of randomness, the band structure is calculated using the supercell method. In Fig.4(a1), the orientation of $\mathbf{M}$ is fixed in $x$-direction, i.e., $M=M_x$; in Fig.4(a2), the orientation of $\mathbf{M}$ uniformly distributes in the whole $4\pi$ solid angle; in Fig.4(a3), $\mathbf{M}$ is located in the $x$-$y$ plane, the orientations of $\mathbf{M}$ are uniformly distributed in $x$-$y$ plane. In these three cases, the chiral edge states do not appear because the average $\mathbf{M}_z$ are zero.
In the bottom panels, average $\mathbf{M}$ has $z$-component. For instance, in Fig.4(b1), the direction of $\mathbf{M}$ is fixed in $z$-direction, i.e., $\mathbf{M}=\mathbf{M}_z$; in Fig.4(b2), $\mathbf{M}$ is uniformly distributed in the upper hemisphere; in Fig.4(b3), $\mathbf{M}$ is a combination of $\mathbf{M}=\mathbf{M}_{xy}$ and $\mathbf{M}=\mathbf{M}_z$. In these three cases, we find $\mathbf{M}_z \ge 0$ for all dopants and the chiral edge state, which is the signature of QAHE, emerges. From Fig.4, we tend to conclude that, as long as $\mathbf{M}_z \ge 0$ for all dopants QAH states would appear. This fact provides an explanation why magnetic disorders does not suppress the QAHE in MTI films. In the following we try to understand the reason why moderate angular randomness of $\mathbf{M}$ favors the formation of QAHE in MTI films. It is known that for 3D MTI films the top and bottom surface states are forbidden in the energy gap induced by $\mathbf{M}_z$, and the edge states can only propagate along the gapless side surfaces. However, since the MTI film is very thin, the energies of the side surface are quantized. As a result the side surface state may not be available for an electron with a given energy. Angular randomness of $\mathbf{M}$ is helpful in suppressing the discreteness of the side energy bands and providing surface states for the electron to propagate for the same energy. Consequently, QAHE in MTI films is substantially improved by magnetic disorders.

Since the magnetic disorders are beneficial to the formation of QAHE in MTI films, we consider a moderate angular randomness of $\mathbf{M}$ by choosing $\Theta=0.46\pi$, and study the effect of magnetic randomness on QAHE in MTI films as we vary the thickness of the thin film. In Fig.5
we plot the average $\rho_{xx}$, the average $\rho_{xy}$ and their fluctuations as a function of film thickness $C_z$ for different magnetic fields $B$.
Comparing Fig.5(a) with Fig.2, we find that with the increase of $C_z$, the average Hall resistance $\rho_{xy}$ is still perfectly quantized. However, the average longitudinal resistance $\rho_{xx}$ is drastically affected by the randomness.
In the presence of random $\mathbf{M}$ with $\Theta=0.46\pi$, $\rho_{xx}$ is suppressed abruptly with the increasing film thickness $C_z$. At $C_z \approx 10a$, $\rho_{xx}$ drops to nearly zero, regardless of the magnetic field strength.
For the clean sample the behavior is the opposite. In Fig.2, where a fixed $\mathbf{M}_z$ is considered, it shows that $\rho_{xx}$ is maximum at $C_z \approx 10a$.
Fig.5(b1) shows that accompanying the nearly zero $\rho_{xx}$, its fluctuation $\Delta\rho_{xx}$ reaches the lowest value as well.
These observations clearly show that, in a MTI film, angular randomness of $\mathbf{M}$ is most effective in suppressing the dissipation of chiral edge states at the film thickness $C_z \approx 10a$.
For $C_z>10a$, $\Delta\rho_{xx}$ increases slowly. Different from $\Delta\rho_{xx}$, the fluctuation of Hall resistance $\Delta\rho_{xy}$ is very small, it is two orders of magnitude smaller than $\Delta\rho_{xx}$, which means that $\rho_{xy}$ is immune to angular randomness of $\mathbf{M}$. In summary, magnetic disorders are beneficial to enhance the quality of edge states of QAHE in MTI films, especially at certain film thickness.

\begin{figure}[tbp]
\includegraphics[width=8.7cm, clip=]{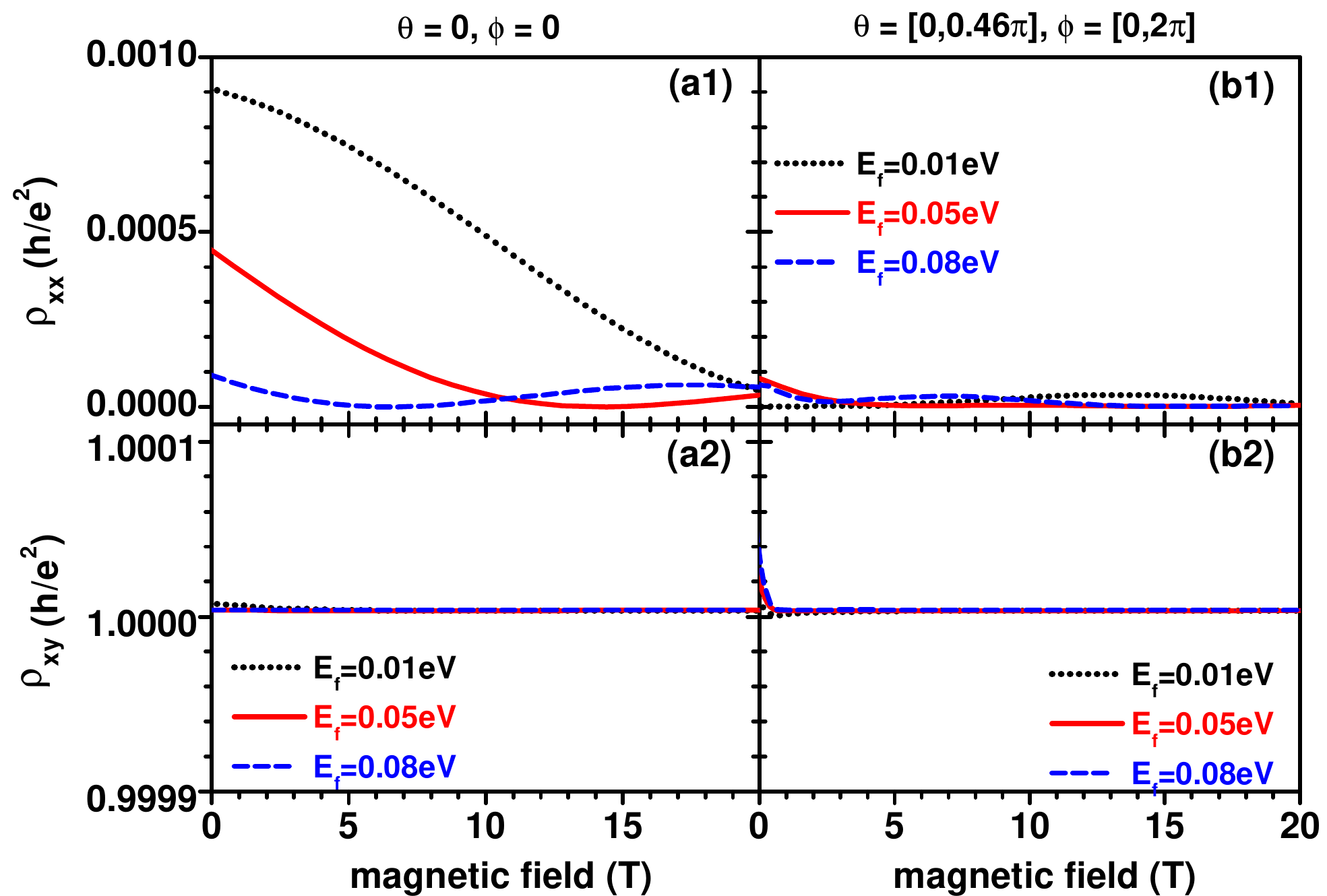}
\caption{ (Color online)
$\rho_{xx}$ and $\rho_{xy}$ vs magnetic field strength with a constant or random exchange field at different Fermi energies.
Left panels: unitary $\mathbf{M}$, $M_z=0.15$eV, $M_{xy}=0$; Right panels: random $\mathbf{M}$, $M_z=0$, $|M_{xy}|=0.15eV$ randomly distributes in the $x$-$y$ plane, i.e., $\theta=0.5\pi$, $\phi=[0,2\pi]$. Averaged over 200 random configurations.}
\end{figure}

To further illustrate the effect of magnetic disorders on QAHE, the QAHE with or without magnetic disorders are compared in Fig.6. The average longitudinal resistance and Hall resistance with constant and random exchange field $\mathbf{M}$ are depicted in Fig.6(a) and Fig.6(b), respectively. Here, the constant exchange field $\mathbf{M}$ is fixed at $\mathbf{M}_z=0.15$eV and $\mathbf{M}_{xy}=0$. Meanwhile, for the case of magnetic disorders, $\mathbf{M}$ is randomly distributed in almost whole upper
hemisphere with $\theta=[0,0.46\pi]$ and $\phi=[0,2\pi]$.
From Fig.6(a) where $\mathbf{M}$ is a constant, we find that the average Hall resistance $\rho_{xy}$ is perfectly quantized, but the average longitudinal resistance $\rho_{xx}$ deviates significantly from zero. On the other hand, Fig.6(b) shows that the deviation in $\rho_{xx}$ is suppressed by magnetic disorders while the perfect $\rho_{xy}$ is maintained.
These findings further confirm that magnetic disorders are useful to eliminate the dissipation of the edge states,
regardless of the strength of external magnetic field.
\begin{figure}
\includegraphics[width=8.7cm, clip=]{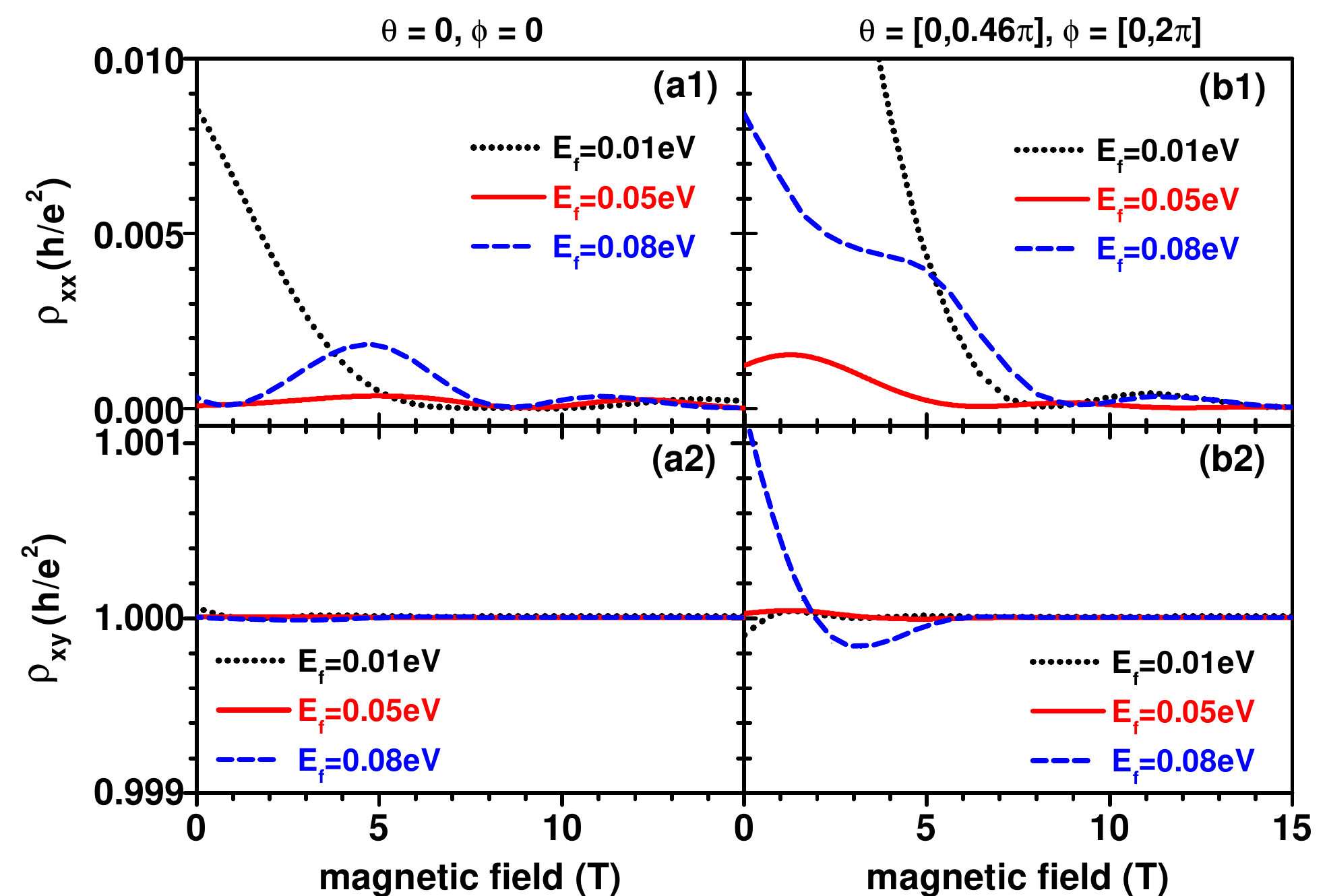}
\caption{ (Color online)
$\rho_{xx}$ and $\rho_{xy}$ vs magnetic field strength with unitary or random exchange field at different Fermi energies. The two-dimensional effective Hamiltonian is adopted in the calculation and other computational parameters are the same as in Fig.6.
Averaged over 500 random configurations.}
\end{figure}

It should be noted that, all conclusions discussed above are only valid for 3D MTI systems with finite thickness, in which edge states can propagate through the gapless side surfaces. For a very thin film, the energy of surface state on side surfaces is discretized so that they may not be available for edge state to propagate. The magnetic disorders tend to overcome the discrete nature of the side surface states, so they are useful in the formation of QAHE in MTI films. On the contrary, in two-dimensional systems, edge states reside on the edges in $x$-$y$ plane. Therefore, any type of random disorders, magnetic or nonmagnatic, would induce the scattering between edge states locating in different edges if the disorder strength is large enough and consequently be destructive to QAHE. To verify this statement, we calculate the longitudinal and Hall resistances using the 2D effective Hamiltonian with magnetic disorders.\cite{Science329.61-64.Yu.2010,PhysicalReviewLetters111.146802.Lu.2013,Phys.Rev.B89.155419.Zhang.2014}
Similar to Fig.6, average $\rho_{xx}$ and $\rho_{xy}$ of the 2D effective model with constant or random exchange field $\mathbf{M}$ are depicted in Fig.7(a) and Fig.7(b), respectively. Comparing Fig.7(a) with Fig.7(b), we see that in the presence of random magnetic disorders, although $\rho_{xy}$ is roughly the same as that of clean sample for two Fermi energies, $\rho_{xx}$ deviate significantly from zero showing strong back scattering. This means that the edge states of 2D systems are heavily damaged by magnetic disorders.
This is totally different from 3D case as shown in Fig.6, in which $\rho_{xx}$ is strongly suppressed and $\rho_{xy}$ is perfectly kept in the presence of magnetic disorders.

\bigskip

\section{conclusion}

In summary, we have studied the influence of magnetic disorders on QAHE in 3D MTI thin films. Magnetic disorders are modeled by random distribution of orientations of the exchange field $\mathbf{M}$. It is found that in the presence of angular randomness of exchange field, QAHE is well kept as long as the $z$-component of $\mathbf{M}$ for all dopants remain positive. Moreover, magnetic disorders are helpful in suppressing the dissipation of the chiral edge states due to the presence of the side surfaces in MTI films. Our results also show that the longitudinal resistance $\rho_{xx}$ relies much on the thickness of the film. At certain film thickness $C_z\approx10a$, magnetic disorders are most effective in protecting the chiral edge states of QAHE. These findings are new features for QAHE in three-dimensional systems, not present in two-dimensional systems.

$${\bf ACKNOWLEDGMENTS}$$
This work is financially supported by the MOST project of China (Grant No. 2016YFA0300603, No. 2015CB921102 and No. 2014CB920903), NNSF project of China (Grant No.11674024, No.11504240, No.11574029, No.11574007, No.11374246), the MOE project of China (No. NCET-13-0048), Research Grant Council (Grant No. 17311116) and the University Grant Council (Contract No.AoE/P-04/08) of the Government of HKSARY,  the grant of BIT (Grant No.20161842028), NSFC of SZU (Grant No.201550 and No.201552).
\\


\end{document}